\input{aipcheck}

\documentclass[
    ,final            
  ]
  {aipproc}

\layoutstyle{8x11double}

\def\siml{{\ \lower-1.2pt\vbox{\hbox{\rlap{$<$}\lower6pt\vbox{\hbox{$\sim$}}}}\ }}
\def\simg{{\ \lower-1.2pt\vbox{\hbox{\rlap{$>$}\lower6pt\vbox{\hbox{$\sim$}}}}\ }}

\def \als {\alpha_{\mathrm{s}}}

\def \m2   {\mu^{2 \epsilon}}
\def\lQ{\Lambda_{\rm QCD}}
\def\als{\alpha_{\rm s}} 
\def\bea{\begin{eqnarray}}
\def\eea{\end{eqnarray}}
\def\be{\begin{equation}}
\def\ee{\end{equation}}

\begin{document}
\title{Heavy Quarkonium at finite temperature}


\author{Nora Brambilla}{
  address={Physik-Department, Technische Universit\"at M\"unchen,
James-Franck-Str. 1, 85748 Garching, Germany}
}

\begin{abstract}
I discuss quarkonium physics at finite temperature in 
the framework of nonrelativistic effective field theories.
\end{abstract}

\maketitle


\section{Quarkonium as a probe of Quark Gluon Plasma Formation}

The study of quarkonium in media has recently undergone crucial developments (see e.g. \cite{Brambilla:2010cs}).
Large datasets from heavy-ion collisions
have recently  become available at RHIC  displaying new features  related
to the quark gluon plasma formation characteristics like the particular
structure of jet quenching and the  very low viscosity to entropy 
ratio. In particular the  quark gluon plasma looks more like a liquid than a 
plasma and the use of perturbative expansion  appears  to be   justified only at 
temperature bigger than the deconfinement one.

The suppression of quarkonium production in
the hot medium remains one of  the cleanest and most relevant
probe of deconfined matter.
However, the use of quarkonium yields as a hot-medium
diagnostic tool has turned out to be quite challenging
for several reasons.

Quarkonium production has already been found to
be suppressed in proton-nucleus collisions by
cold-nuclear-matter effects, which themselves require
dedicated experimental and theoretical attention. Recombination effects
may  play an additional role and  thus transport properties may become 
relevant to  be considered. Finally,  the heavy quark-antiquark interaction
at finite temperature $T$ has to be obtained from QCD  \cite{Brambilla:2004wf,Brambilla:2010cs}.

\section{The Heavy Quark Interaction at Finite T}

Since the publication of the famous 
paper \cite{Matsui:1986dk} of Matsui and Satz arguing that color screening 
in a deconfined QCD medium could destroy all $Q\overline Q$ bound states
at sufficiently high temperatures, 
there has been considerable interest in studying the interaction between 
a heavy quark and a heavy antiquark  in hot media. The expectation was 
that the linearly confining potential of zero temperature would get replaced by a 
Debye-screened potential at high temperatures  \cite{Matsui:1986dk}.
However up to very recent times no proper tool to define and calculate the quarkonium potential
at finite T was developed. Most of the investigations have been 
performed with phenomenological potentials  whose behaviour was 
inspired by lattice calculations of the free energy.
Color screening is indeed studied on the lattice by 
calculating the spatial correlation functions of a static quark and
antiquark in a color-singlet state which propagates in Euclidean time 
from $\tau=0$ to $\tau=1/T$, where $T$ is the temperature 
(see e.g.  \cite{Brambilla:2010cs,Satz:2000bn,Philipsen:2008qx}
 for reviews). 
Lattice calculations of this quantity with dynamical quarks have been
also reported. 
The logarithm of the singlet
correlation function is called the singlet free energy,
  In the zero-temperature limit the
singlet free energy coincides with the zero-temperature potential.
Moreover  at sufficiently short  $Q\bar{Q}$  distances, the singlet free energy is
temperature independent and equal to the zero-temperature potential,  while for large distance 
it shows a flattening behaviour that is thought to be related to the screening effect.
The range of interaction decreases with increasing temperature.  For 
temperatures above the transition temperature, $T_c$, the heavy-quark 
interaction range becomes comparable to the charmonium radius. Based on this general observation, 
one would expect that the charmonium
states, as well as the excited bottomonium states, do not remain bound at
temperatures just above the
deconfinement transition, and this  referred to as 
quarkonium {\it dissociation} or quarkonium {\it melting}.
The free energy is extracted on the lattice typically from a calculation of 
the quark-antiquark Polyakov loop correlator. There are singlet and octet channels that are  
gauge dependent and one can define also an average gauge-independent free energy.
The three different channels show a different dependency on the $Q\bar{Q}$ separation 
and thus lead to different binding energies when used as phenomenological potentials
in the Schr\"odinger equation \cite{Philipsen:2008qx}. There is a huge literature using the singlet free energy 
or the corresponding internal energy to calculate quarkonium energies at finite T 
and recostructing the lattice meson correlation functions to understand which one fits better.

It is therefore very important to find a theoretical framework that can give 
us the definition of what is the $Q\bar{Q}$ potential at finite T and a calculation tool.
This is realized by constructing an appropriate effective field theory (EFT).

For observables only sensitive to gluons and light quarks,
a very successfull EFT called Hard Thermal Loop (HTL)
effective theory has been derived  in the past \cite{Braaten:1989mz}
by integrating out the hardest momenta  propotional to $T$
from the dynamics.  However,   considering also heavy quarkonium in the hot 
QCD medium,  one has to consider in addition to the thermodynamical scales 
in $T$ also the scales of the nonrelativistic bound state and the situation 
becomes more complicate. 

In the last few years years, there has been a remarkable 
progress in addressing the problem of outlining an  EFT framework   for quarkonium at finite temperature and 
in rigorously defining 
the quarkonium potential.
In \cite{Laine:2006ns,Laine:2007qy}, the static potential was calculated 
in the regime $T \gg 1/r \simg m_D$, where $m_D$ is the Debye mass and 
$r$ the quark-antiquark 
distance, by performing an analytical continuation of the 
Euclidean Wilson loop
to real time. The calculation was done in 
the weak-coupling resummed perturbation
theory. The imaginary part of the gluon self energy gives an imaginary part to
the static potential and hence a thermal width to
the quark-antiquark bound state. In the same framework, the dilepton
production rate for charmonium and bottomonium was calculated in 
\cite{Laine:2007gj,Burnier:2007qm}.
In \cite{Beraudo:2007ky}, static particles in real-time formalism were 
considered  and the potential for distances $1/r \sim m_D$ was derived 
for a hot QED plasma. 
The real part of the static potential was found to agree with the
singlet free energy and the damping factor with the one found in 
\cite{Laine:2006ns}.
In \cite{Escobedo:2008sy}, a study of bound states in a hot QED 
plasma was performed in a non-relativistic EFT framework. 
In particular, the hydrogen atom was studied for temperatures ranging from 
$T\ll m\alpha^2$ to $T\sim m$, where the imaginary part of the
potential becomes larger than the real part and the hydrogen ceases to exist. 
The same study has been extended to muonic hydrogen in 
 \cite{Escobedo:2010tu}, providing a method to estimate  the effects of a 
finite charm quark mass on the dissociation temperature of bottomonium.
In the next sections we report our work in the construction of a 
an EFT description of heavy quarkonium at finite T. 
We study the real-time evolution of a static quark-antiquark pair in a medium of gluons and light quarks at finite temperature. For temperatures ranging from values larger to smaller than the inverse distance of the quark and antiquark, 
and at short distances, we derive the potential between the two static sources, 
and calculate their energy and thermal decay width. 
We will see in particular 
that the  EFT enables us to give both a proper definition of the heavy quarkonium 
potential inside a hot medium and solid calculation tools to obtain it (at least in the weak 
coupling situation).

\section{An EFT  framework}

An EFT framework in real time and weak coupling for quarkonium at finite
temperature was developed in \cite{Brambilla:2008cx}   working in real time and 
in the regime 
of small coupling $g$, so that $g T \ll T$  and  $v \sim \als$, which is expected to be valid 
for tightly bound states: $\Upsilon(1S)$, $J/\psi$, ...~.

 Quarkonium in a medium is characterized by different energy and momentum scales.
   There are the scales of the non-relativistic bound state:
($v$ is the relative heavy-quark velocity):
$m$, the heavy quark mass, $mv$, the scale of the typical inverse 
distance between the heavy quark and antiquark, $mv^2$, the scale of the 
 typical binding energy or potential and lower energy scale.  Furthermore 
 there are the   thermodynamical scales: the temperature $T$, the inverse of the screening 
length of the chromoelectric interactions, i.e. the Debye mass $m_D$  ($\sim gT$ in the perturbative regime)
 and lower scales, which  we will neglect in the following.

If these  scales are hierarchically ordered,  then we may expand physical observables in the 
ratio of such scales. If we separate explicitly the contributions from the different scales
at the Lagrangian level this amounts to substituting QCD with a hierarchy of EFTs, which are equivalent 
to QCD order by order in the expansion parameters.
At zero temperature the EFTs
that follow from QCD by integrating out the scales $m$ and $mv$ are called respectively 
Non-relativistic QCD (NRQCD) \cite{Caswell:1985ui}
 and potential NRQCD (pNRQCD)    \cite{Pineda:1997bj,Brambilla:1999xf ,Brambilla:2004jw}.

We assume that the temperature is high enough that $T \gg gT \sim m_D$ holds 
but also that it is low enough for $T \ll m$ and $1/r \sim mv \simg m_D$ to be satisfied, 
because for higher temperature  the bound state ceases to exist.
Under these conditions some possibilities are in order. If $T$ is the next relevant scale after 
$m$, then integrating out $T$ from NRQCD leads to an EFT that we may name NRQCD$_{\rm HTL}$, because 
it contains the hard thermal loop (HTL) Lagrangian
 \cite{Braaten:1989mz}.
Subsequently integrating out 
the scale $mv$ from NRQCD$_{\rm HTL}$ leads to a thermal version of pNRQCD that we may call 
pNRQCD$_{\rm HTL}$. If the next relevant scale after $m$ is $mv$, 
then integrating out $mv$ from NRQCD leads to pNRQCD. If the temperature is larger than $mv^2$, 
then the temperature may be integrated out from pNRQCD leading to a new version of pNRQCD$_{\rm HTL}$  \cite{Vairo:2009ih}.
 Note that, as long as the temperature is smaller than the scale being 
integrated out, the matching leading to the EFT may be performed putting the
 temperature to zero.

The derived potential $V$  describes the real-time evolution of a quarkonium state
in a thermal medium. At leading order, the evolution is governed 
by a Schr\"o\-din\-ger equation. In an EFT framework,
the potential follows naturally from integrating out all  contributions coming from modes
with energy and momentum larger than the binding energy.
For $T < V$ the potential is simply the Coulomb potential. Thermal corrections 
affect the energy and induce a thermal width to the quarkonium state; these 
may be relevant to describe the in medium modifications of quarkonium at low temperatures.
For $T >V$ the potential gets thermal contributions, which are both real and imaginary.

\section{Results of the EFT description}

General findings in this picture  are:
\begin{itemize}
\item{}  The thermal part of the potential has a real and  an imaginary part. 
The imaginary part of the potential smears out the bound state 
peaks of the quarkonium spectral function, leading to their dissolution prior
to the onset of Debye screening in 
the real part of the potential (see, e.g. the 
discussion in \cite{Laine:2008cf}).
So quarkonium dissociation appears to  be a consequence of the 
appearance of a thermal decay width rather than being due to the color screening of
the real part of the potential; this follows from the observation that the
thermal decay width becomes as large as the binding energy at a temperature 
at which color screening may not yet have set in.
\item{} Two mechanisms contribute to the thermal decay width: the imaginary part of the gluon self energy  induced by the Landau-damping phenomenon (existing also in QED)  
\cite{Laine:2006ns}  and the quark-antiquark color singlet to color 
octet thermal break up (a new effect, specific of QCD)  \cite{Brambilla:2008cx}.
 Parametrically, the first mechanism dominates for temperatures 
such that the Debye mass $m_D$ is larger than the binding energy, while the latter 
dominates for temperatures such that $m_D$  is smaller than the binding energy.
\item{} The obtained singlet thermal potential, $V$, is neither the color-singlet quark-antiquark free energy  nor the internal energy. It has an imaginary part and may contain divergences
that eventually cancel in physical observables \cite{Brambilla:2008cx}.
\item{} Temperature effects can be other than screening, typically they  may appear as
power law corrections or a logarithmic dependence \cite{Brambilla:2008cx,Escobedo:2008sy}.
\item{} The dissociation temperature goes parametrically as  $\pi T_{\rm melting} \sim m g^{4\over 3}$ \cite{Escobedo:2008sy,Laine:2008cf}.
\end{itemize}

\section{The Free Energy in the EFT approach}

In \cite{Brambilla:2010xn,Burnier:2009bk} the Polyakov loop  and the correlator of two Polyakov loops 
at finite temperature  has ben calculated at next-to-next-to-leading order in the weak coupling 
regime and at quark-antiquark distances shorter than the inverse of the temperature and for 
Debye mass larger than the Coulomb potential.  
The calculation has been performed also the in EFT framework 
\cite{Brambilla:2010xn} and a relation between the Polyakov loop correlator and the  singlet and octet  quark-antiquark correlator      
\cite{Brambilla:2008cx}
has been established in this setup.

\section{Calculation of Bottomonium  properties at LHC}

The EFT  provides a clear definition of the potential and a  coherent and systematical setup
to calculate the masses and widths of the lowest quarkonium resonances at finite temperature.
 In \cite{Brambilla:2010vq} 
heavy quarkonium energy levels and decay widths in a quark-gluon plasma, 
below the melting temperature at a
temperature $T$ and screening mass $m_D$ satisfying the hierarchy  
\begin{equation}
m \gg m\als \gg \pi T \gg m\als^2 \gg m_D,
\label{hierarchy}
\end{equation}
have been calculated  at order $m \als^5$.
This implies that $mg^3 \gg T \gg mg^4$ and that 
$\pi T$ is lower than $\pi T_{\rm melting}$, i.e. quarkonium exists in the plasma.
We will further assume that all these scales are larger than $\lQ$ and that 
a weak-coupling expansion is possible for all of them.
Finally, in order to produce an expression for the spectrum that is accurate up to order 
$m\als^5$, we will assume that $[m_D/(m\als^2)]^4 \ll g$. 

This  situation may be  relevant for bottomonium $1S$   states ($\Upsilon(1S)$, $\eta_b$) at the LHC,
for which 
it may hold $m_b \approx 5 \; \hbox{GeV} \; > m_b\als \approx 1.5 \; \hbox{GeV} \; >  \pi T \approx 1 \; 
\hbox{GeV} \; >  m\als^2 \approx 0.5  \; \hbox{GeV} \; \simg  m_D$.

We work in the  real-time formalism,  that allows us to develop  a treatment 
of the quarkonium in the thermal bath very similar to the EFT framework developed 
for zero temperature \cite{Brambilla:2004jw}.
We integrate out the mass and the relative momentum arriving at pNRQCD.
According to the hierarchy \eqref{hierarchy}, both these scales are larger 
than $T$, which, therefore, may be set to zero in the matching to the EFT.
As a consequence, the Lagrangians of NRQCD and pNRQCD are the same as at zero temperature.

Integrating out $T$ from pNRQCD modifies pNRQCD into hard-thermal loop (HTL) pNRQCD, 
pNRQCD$_{\rm HTL}$, \cite{Brambilla:2008cx,Vairo:2009ih}. With respect to pNRQCD, 
the pNRQCD$_{\rm HTL}$  Lagrangian gets relevant modifications in two parts.
First, the Yang--Mills Lagrangian gets an additional HTL part \cite{Braaten:1989mz}.
This, for instance, modifies the longitudinal gluon propagator in Coulomb gauge into ($k^2 \equiv {\bf k}^2$)
\begin{equation}
\frac{i}{k ^2}
\to 
\frac{i}{k^2+m_D^2\left(1-\displaystyle\frac{k_0}{2k}\ln\frac{k_0+k\pm i\eta}{k_0-k\pm i\eta}\right)},
\end{equation}
where ``$+$'' identifies the retarded and ``$-$'' the advanced propagator.
Second, the potentials get in addition to the Coulomb potential, which 
is the potential inherited from pNRQCD, a thermal part, that we call $\delta V$. 

\subsection{Potential, energy and decay width}
In the following, we will provide the thermal corrections to the color-singlet 
quark-antiquark potential, the thermal corrections to the spectrum and the thermal decay width, 
aiming at a precision of the order of $m\als^5$.

\begin{figure}[ht]
\resizebox{12pc}{!}{\includegraphics{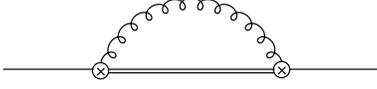}}
\label{fig1}
\caption{The single line stands for a quark-antiquark color-singlet propagator,
the double line for a quark-antiquark color-octet propagator and the circle with a cross for 
a chromoelectric dipole interaction.}
\end{figure}

\subsubsection{Integrating out the scale $T$}
The relevant diagram contributing to the potential is shown in Fig.~\ref{fig1}. It reads
\begin{eqnarray}
 &&
- i g^2 \, \frac{4}{3} \, \frac{r^i}{D-1}
\mu^{4-D} \int \frac{d^Dk}{(2\pi)^D}
\frac{i}{E - h_o -k_0 +i\eta}   \nonumber  \\
&& \times \left[k_0^2 \, D_{ii}(k_0,k) +  k^2 \, D_{00}(k_0,k)
\right]r^i, 
\end{eqnarray}
where $D_{\mu\nu}$ stands for the gluon propagator, $h_o = {\bf p}^2/m + \als/(6r)$ is the octet 
Hamiltonian and the loop integral has been regularized in dimensional regularization ($D = 4 + \epsilon$ 
and $\mu$ is the subtraction point). In the loop integral, we integrate over 
the momentum region $k_0\sim T$ and $k \sim T$.
Since $T\gg(E-h_o)$, we may expand 
\begin{eqnarray}
&& \frac{i}{E-h_o-k_0+i\eta} = 
 \frac{i}{-k_0+i\eta}-i\frac{E-h_o}{(-k_0+i\eta)^2} \nonumber  \\
&& +i\frac{(E-h_o)^2}{(-k_0+i\eta)^3}-i\frac{(E-h_o)^3}{(-k_0+i\eta)^4}+\dots\,.
\end{eqnarray}


The real part of the thermal correction to the color-singlet potential reads
\begin{eqnarray}
\!\!\!\!\!\!\!\!\!\!  && {\rm Re}~ \delta V_s(r) =  \frac{4\pi}{9}  \als^2 \, r \, T^2 + \frac{8\pi}{9m} \als \,T^2
+\frac{4\als I_T}{3\pi}   \nonumber \\
&& \left[-\frac{9}{8}\frac{\als^3}{r}- 
\frac{17}{3}\frac{\als^2}{mr^2} 
        +\frac{4}{9}\frac{\pi\als}{m^2}\delta^3({\bf r})+ \frac{\als}{m^2}\left\{\nabla^2_{\bf r},\frac{1}{r}\right\}\right]
\nonumber\\
& &  - 2 \zeta(3) \frac{\als}{\pi} \, r^2 \, T \,m_D^2 + \frac{8}{3} \zeta(3) \als^2 \, r^2 \, T^3 , 
\label{realV}\\
&& I_T   =  \frac{2}{\epsilon}+\ln\frac{T^2}{\mu^2}-\gamma_E+\ln( 4\pi)-\frac{5}{3},
\\
&& m_D^2 = g^2T^2\left(1 + \frac{n_f}{6} \right),
\label{mD}
\end{eqnarray}
where the first line of Eq. \eqref{realV} is of order  $ g^2r^2T^3\times E/T$ and comes from the second term 
in Eq.(2);
the second one is of order $g^2r^2T^3\times (E/T)^3$ and come from the fourth term in Eq.(4), 
and the third one,  is of order $g^2r^2T^3\times ({m_D}/{T})^2$, comes from a self energy insertion 
inside the gluon propagator in  diagram Fig.~\ref{fig1}; $n_f$ is the number of light quarks.

The imaginary part of the color-singlet potential, 
comes from the imaginary part of the self energy graph mentioned above
 and  reads
\begin{eqnarray}
&&  {\rm Im}~\delta V_s(r)   \!\!\! =  \!\!\!
\frac{2}{9} \als \, r^2 \, T \,m_D^2\, \left( 
-\frac{2}{ \epsilon} + \gamma_E + \ln\pi 
- \ln\frac{T^2}{\mu^2} + \frac{2}{3} \right. 
\nonumber\\
&& \left. - 4 \ln 2 - 2 \frac{\zeta^\prime(2)}{\zeta(2)} \right)
+ \frac{16\pi}{9} \ln 2 \,  \als^2\, r^2 \, T^3.
\end{eqnarray}
This contribution, which may be traced back to the Landau-damping phenomenon, 
is of order $g^2r^2T^3\times \left({m_D}/{T}\right)^2$.  

Evaluating ${\rm Re}~\delta V_s(r)$ and ${\rm Im}~\delta V_s(r)$ on a quarkonium state 
with quantum numbers $n$ and $l$, we obtain the thermal correction to the energy, 
$\delta E_{n,l}^{(T)}$, and the thermal width, $\Gamma_{n,l}^{(T)}$, coming from  the scale $T$:
\begin{eqnarray}
&& \delta E_{n,l}^{(T)}  = 
\frac{2\pi}{9}\,\als^2 \,T^2 a_0 \left[3n^2-l(l+1)\right]
+\frac{8\pi}{9m} \als\, T^2
\nonumber\\
&&
+\frac{E_n { I_T}  \als^3}{3\pi}\left\{
-\frac{32}{27}\frac{\delta_{l0}}{n} 
+\frac{200}{3}\frac{1}{n (2l+1)}
-\frac{16}{3}\frac{1}{n^2}
+\frac{27}{4}
\right\}
\nonumber\\
&& 
+\left(- \zeta(3) \frac{\als}{\pi}  \, T \,m_D^2
+ \frac{4}{3} \zeta(3) \als^2 \, T^3\right) \nonumber \\
&& a_0^2n^2 \left[5n^2+1-3l(l+1)\right], 
\label{ET}\\
&& \Gamma_{n,l}^{(T)}  = 
\left[ - \frac{2}{9} \als T m_D^2
\left( -\frac{2}{\epsilon} + \gamma_E + \ln\pi 
- \ln\frac{T^2}{\mu^2}+ \right. \right. \nonumber \\
&& \left.  \frac{2}{3} - 4 \ln 2 - 2 \frac{\zeta^\prime(2)}{\zeta(2)} \right) 
\nonumber\\
&& \left.
-\frac{16\pi}{9} \ln 2 \,  \als^2\, T^3 \right] {a_0^2n^2}\left[5n^2+1-3l(l+1)\right], 
\label{GT}
\end{eqnarray}
where $\displaystyle E_n=-\frac{1}{m a_0^2 n^2} = -\frac{4 m\als^2}{9n^2}$ and $\displaystyle a_0 = \frac{3}{2m\als}$.

\subsubsection{Integrating out the scale $E$}
The diagram shown in Fig.~\ref{fig1} also carries contributions coming from the energy scale $E$.
They may be best evaluated in pNRQCD$_{\rm HTL}$ by integrating 
over the momentum region $k_0\sim E$ and $k \sim E$ and using HTL gluon propagators.
Since $k\sim E \ll T$, we may expand the Bose--Einstein distribution
\begin{equation}
n_{\rm B}(k)=\frac{T}{k}-\frac{1}{2}+\frac{k}{12 \, T} + \dots \;;
\label{bose}
\end{equation}
moreover, since  $k\sim E\gg m_D$, the HTL propagators can be expanded in $m_D^2/E^2\ll 1$.

The momentum region $k_0\sim E$ and $k \sim E$ is characterized by two possible 
momentum sub-regions. This can be understood by considering the integral  
\begin{eqnarray}
&& \int\!\!\frac{d^{D-1}k}{(2\pi)^{D-1}}\int_0^\infty\!\!\frac{dk_0}{2\pi} 
\frac{1}{k_0^2 -k^2 - m_D^2 + i\eta}
\nonumber \\
&& \left(\frac{1}{E-h_o-k_0+i\eta}+\frac{1}{E-h_o+k_0+i\eta}\right). 
\end{eqnarray}
For $k_0\sim E$ and $k \sim E$, it exhibits an off-shell sub-region, $k_0-k \sim  E$, 
and a collinear sub-region, $k_0-k \sim  m_D^2/E$.
Note that, according to \eqref{hierarchy}, the collinear scale satisfies 
$mg^4 \gg  m_D^2/E \gg mg^6$, i.e. it is smaller than $m_D$ by a factor of $m_D/E\ll 1$ 
but still larger than the non-perturbative scale $g^2 T$ by a factor $T/E \gg 1$.

The thermal correction to the energy, $\delta E_{n,l}^{(E)}$, coming from  the scale $E$, reads
\begin{equation}
\delta E_{n,l}^{(E)}  = -\frac{2\pi}{9} \als \, Tm_D^2 \, a_0^2n^2 \left[5n^2+1-3l(l+1)\right].
\label{EE}
\end{equation}
We note the complete cancellation of the vacuum contribution (which includes the Bethe logarithm) 
against the thermal contribution originating from the ``$-1/2$'' term in the expansion of the Bose--Einstein 
distribution (see Eq. \eqref{bose}).

The thermal width, $\Gamma_{n,l}^{(E)}$, coming from  the scale $E$, reads
\begin{eqnarray}
&& \Gamma_{n,l}^{(E)}  =
4\als^3T-\frac{64}{9m}\als TE_n+\frac{32}{3}\als^2T\frac{1}{mn^2a_0}
\nonumber\\
&&
+\frac{2E_n\als^3}{3}
\left\{
-\frac{32}{27}\frac{\delta_{l0}}{n} 
+\frac{200}{3}\frac{1}{n (2l+1)}
-\frac{16}{3}\frac{1}{n^2}
+\frac{27}{4}
\right\}
\nonumber\\
&&
-\frac{2}{9} \als Tm_D^2 \left(\frac{2}{ \epsilon}
+\ln\frac{E_1^2}{\mu^2}+\gamma_E-\frac{11}{3}-\ln\pi+\ln4\right)
\nonumber \\
&&  a_0^2n^2 \left[5n^2+1-3l(l+1)\right]
+\frac{128 Tm_D^2}{81}\frac{\als^3}{E_n^2}\,I_{n,l} \,,
\label{GE}
\end{eqnarray}
where $I_{1,0}=-0.49673$, $I_{2,0}=0.64070, \; \dots\;$~.
The leading contribution is given by the first three terms, which are of order  $\als^3 T$.
This contribution to the thermal width is generated by the possible break up of a quark-antiquark color-singlet state 
into an unbound quark-antiquark color-octet state: a process that is kinematically allowed only in a medium.
The singlet to octet break up is a different phenomenon with respect 
to the Landau damping. In the situation $E \gg m_D$, the first dominates over the second 
by a factor $(m\als^2/m_D)^2$.

\subsubsection{Integrating out the scale $m_D$}
The diagram shown in Fig.~\ref{fig1} also carries contributions coming from the energy scale $m_D$.
These contributions are suppressed with respect to the other terms calculated.

\subsection{Cancellation of divergences}
The thermal corrections to the spectrum and the thermal decay width develop divergences at the different energy scales. 
These are artifacts of the scale separations and cancel in the final (physical) results.

Concerning the thermal decay width, the divergence at the scale  $m\als^2$  in Eq. \eqref{GE}, 
which is of ultraviolet (UV) origin, cancels against the infrared (IR) 
divergence  at the scale  $T$ in Eq. \eqref{GT}.

In the spectrum, the pattern of divergences is more complicated and is summarized in Tab.~\ref{table}.
The table may be read vertically or horizontally. 
If read vertically, it shows a typical EFT cancellation mechanism:
at the scale $m\als$ we have non-thermal IR divergences in the potentials, 
these cancel against non-thermal UV divergences at the scale $m\als^2$ \cite{Brambilla:1999qa}, 
the non-thermal contribution at the scale $T$ is scaleless and vanishes in dimensional regularization; 
thermal IR divergences at the scale $T$ cancel against thermal UV divergences at the scale $m\als^2$. 
If read horizontally, it shows a cancellation mechanism that is familiar in thermal field theory: 
at the scale $T$, thermal IR divergences cancel against non-thermal IR divergences while non-thermal UV 
divergences cancel against IR divergences that appear in the potentials at the scale $m\als$; 
at the scale $m\als^2$ UV thermal divergences cancel against UV non-thermal divergences. 
The cancellation between non-thermal and thermal parts 
is possible because the latter may carry tempe\-ra\-ture independent terms (see, for instance, 
the ``$-1/2$'' term in Eq. \eqref{bose}). 
Note that both at the scales $T$ and $m\als^2$, the spectrum is finite.

\begin{center}
\begin{table}
\begin{tabular}{c|c|c}
Scale  & Vacuum & Thermal \\
\hline 
& & \\
$m\als$ &$\displaystyle \sim m\als^5\frac{1}{\epsilon_{\rm IR}}$ 
& \put(15,-5){\resizebox{2pc}{!}{
\includegraphics{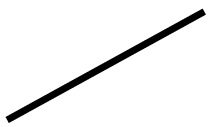}}} \\
& & \\
$T$& $\displaystyle \sim m\als^5\left(\frac{1}{\epsilon_{\rm IR}} -\frac{1}{\epsilon_{\rm UV}} \right)$ 
&$\displaystyle \sim - m\als^5\frac{1}{\epsilon_{\rm IR}}$ \\
& & \\
$m\als^2$ & $\displaystyle \sim -m\als^5\frac{1}{\epsilon_{\rm UV}}$ 
& $\displaystyle \sim m\als^5\frac{1}{\epsilon_{\rm UV}}$\\
\end{tabular}
\caption{The pattern of IR and UV divergences in the quarkonium spectrum at different energy 
scales. The final result is finite.}
\label{table}
\end{table}
\end{center}

\subsection{Results}
For a quarkonium state that satisfies the hierarchy specified in Eq. \eqref{hierarchy} 
and in the following discussion, 
the complete thermal contribution to the spectrum up to ${\cal O}(m\als^5)$ 
is obtained by summing Eqs. \eqref{ET} and \eqref{EE} and subtracting 
from the latter the zero-temperature part, this gives 
\begin{eqnarray}
&& \delta E_{n,l}^{(\mathrm{thermal})} =
\frac{2\pi}{9} \als^2 \,T^2 a_0 \left[3n^2-l(l+1) + \frac{8}{3}\right] 
\nonumber\\
&&
+\frac{E_n\als^3}{3\pi}\left[\log\left(\frac{2\pi T}{E_1}\right)^2-2\gamma_E\right]
\\
&& \times
\left\{
-\frac{32}{27}\frac{\delta_{l0}}{n} 
+\frac{200}{3}\frac{1}{n (2l+1)}
-\frac{16}{3}\frac{1}{n^2}
+\frac{27}{4}
\right\}
\nonumber\\
&&
+\frac{128E_n\als^3}{81\pi}L_{n,l}
+ a_0^2n^2 \left[5n^2+1-3l(l+1)\right]
\nonumber \\
&& \left\{- \left[\frac{1}{\pi} \zeta(3)+\frac{2\pi}{9} \right]   \als  \, T \,m_D^2
\right.
+ \left. 
\frac{4}{3} \zeta(3)\, \als^2 \, T^3\right\}, \nonumber 
\end{eqnarray}
where $L_{n,l}$ are the QCD Bethe logarithms: $L_{1,0}=-81.5379$, 
$L_{2,0}=-37.6710, \; \dots\;$ \cite{Kniehl:2002br}.

For a quarkonium state that satisfies the hierarchy specified in Eq. \eqref{hierarchy} 
and in the following discussion, 
the complete thermal width up to ${\cal O}(m\als^5)$ is obtained by summing 
Eqs.~\eqref{GT} and \eqref{GE}, this gives
\begin{eqnarray}
&& \Gamma_{n,l}^{(\mathrm{thermal})} =
\left( 4 + \frac{832}{81}\frac{1}{n^2} \right)\als^3T
\nonumber\\
&&
+\frac{2E_n\als^3}{3}
\left\{
-\frac{32}{27}\frac{\delta_{l0}}{n} 
+\frac{200}{3}\frac{1}{n (2l+1)}
-\frac{16}{3}\frac{1}{n^2}
+\frac{27}{4}
\right\}
\nonumber\\
&&
-\left[\frac{2}{9} \als T m_D^2
\left(\ln\frac{E_1^2}{T^2}+ 2\gamma_E -3 -\log 4- 2 \frac{\zeta^\prime(2)}{\zeta(2)} \right)
\right.
\nonumber \\
&& \left.
+\frac{16\pi}{9} \ln 2  \,  \als^2\, T^3 \right] 
 a_0^2n^2\left[5n^2+1-3l(l+1)\right]
\nonumber\\
&& + \frac{32}{9}\als\, Tm_D^2\,a_0^2n^4 \,I_{n,l}.
\end{eqnarray}

As a qualitative summary, we observe that, at leading order, the quarkonium masses 
increase quadratically with $T$,
which implies the same functional increase in the
energy of the leptons and photons produced in the electromagnetic decays.
Electromagnetic decays occur at short distances $\sim 1/m \ll 1/T$, 
hence the standard NRQCD factorization formulas hold. At leading order, 
all the temperature dependence is encoded in the wave function
at the origin. The leading temperature correction to it can be read from the 
potential and is of order $n^4 T^2/(m^2\als)$. 
Hence, a  quadratic dependence on the temperature should
be observed in the frequency of produced leptons or photons.
Finally, at leading order, a decay width linear with temperature is developed.
The mechanism underlying this decay width is the color-singlet to color-octet 
thermal break-up, which implies a tendency of the quarkonium to decay 
into a continuum of color-octet states.



\begin{theacknowledgments}
We acknowledge financial support from the DFG cluster of excellence ``Origin and structure of the universe''
({http://www.universe-cluster.de}).
\end{theacknowledgments}



\bibliographystyle{aipproc}   


\IfFileExists{\jobname.bbl}{}
 {\typeout{}
  \typeout{******************************************}
  \typeout{** Please run "bibtex \jobname" to optain}
  \typeout{** the bibliography and then re-run LaTeX}
  \typeout{** twice to fix the references!}
  \typeout{******************************************}
  \typeout{}
 }

\end{document}